\documentclass[a4paper,11pt]{article}

\usepackage{pos}
\usepackage{relsize}

\usepackage[acronyms, nohypertypes={acronym}, nopostdot, style=super, nonumberlist, toc]{glossaries}
\glsenableentrycount
\setacronymstyle{long-sc-short}

\newcommand\ac[1]{\gls{#1}}

\newacronym{WF}{wf}{Wilson-Fisher}
\newacronym{AF}{af}{asymptotically free}
\newacronym{RG}{rg}{renormalization group}
\newacronym{WZW}{wzw}{Wess-Zumino-Witten}
\newacronym[longplural={conformal field theories}]{CFT}{cft}{conformal field theory}
\newacronym[longplural={lattice field theories}]{LFT}{lft}{lattice field theory}
\newacronym[longplural={effective field theories}]{EFT}{eft}{effective field theory}
\newacronym[longplural={quantum field theories}]{QFT}{qft}{quantum field theory}
\newacronym{LEC}{lec}{low-energy constant}
\newacronym{QCD}{qcd}{quantum chromodynamics}
\newacronym{MC}{mc}{Monte Carlo}
\newacronym{IR}{ir}{infrared}
\newacronym{UV}{uv}{ultraviolet}
\newacronym{SNR}{snr}{signal-to-noise ratio}
\newacronym{NLSM}{nl$\sigma$m}{nonlinear sigma model}
\newacronym{PCM}{pcm}{principal chiral model}
\newacronym{CSA}{csa}{Cartan subalgebra}
\newacronym{SSB}{ssb}{spontaneous symmetry breaking}
\newacronym{DOF}{dof}{degrees of freedom}
\newacronym{DMRG}{dmrg}{densiy matrix renormalization group}
\newacronym{YM}{ym}{Yang-Mills}
\newacronym{QLM}{qlm}{quantum link model}
\newacronym{KG}{kg}{Kogut-Susskind}
\newacronym{SPT}{spt}{symmetry-protected topological}
\newacronym{GW}{gw}{Ginsparg-Wilson}
\newacronym{FK}{fk}{Fidkowski-Kitaev}
\newacronym{CS}{cs}{Chern-Simons}
\newacronym{APS}{aps}{Atiyah-Patodi-Singer}
\newacronym{PV}{pv}{Pauli-Villars}
\newacronym{PBC}{pbc}{periodic boundary conditions}
\newacronym{OBC}{obc}{open boundary conditions}
\newacronym{ABC}{abc}{antiperiodic boundary conditions}
\newacronym{KD}{kd}{Kähler-Dirac}
\newacronym{SMG}{smg}{symmetric mass generation}
\newacronym{GB}{cgb}{Chern-Gauss-Bonnet}
\newacronym{WY}{wy}{Witten-Yonekura}
\newacronym{RKD}{rkd}{reduced Kähler-Dirac}
\newacronym{AS}{as}{Atiyah-Singer}
\newacronym{BC}{bc}{boundary conditions}
\newacronym{EWBG}{EWBG}{electroweak baryogenesis}
\newacronym{INT}{int}{Institute for Nuclear Theory}
\newacronym{NN}{nn}{Nielsen-Ninomiya}

\usepackage{mathtools}
\usepackage{amsmath,amsbsy,amsfonts}
\usepackage{ninecolors}
\usepackage{microtype}
\usepackage{xparse}

\usepackage{physics}
\usepackage{physics2}
\usephysicsmodule{ab,ab.braket}

\usepackage{graphicx}
\usepackage{amsfonts}
\usepackage{tocvsec2}
\usepackage{slashed}
\usepackage{cancel}
\usepackage{comment}

\usepackage{graphics}
\graphicspath{{./fig/}}
\usepackage{tikz}
\usetikzlibrary{3d}

\usepackage[section]{placeins}

\usepackage[normalem]{ulem}
\usepackage{minibox}
\usepackage{orcidlink}

\usepackage{enumitem}

\usepackage{hyperref}
\hypersetup{unicode, breaklinks, colorlinks=true, allcolors=azure3}
\usepackage[capitalise]{cleveref}
\crefname{section}{Sec.}{Secs.}

\usepackage{xcolor}
\usepackage{colortbl}
\usepackage{tabularx}

\usepackage{makecell}
\newcommand\TopRule{\Xhline{0.08em}}

\newcommand\MidRule{\Xhline{0.03em}}
\newcommand\BotRule{\Xhline{0.08em}}

\newcommand\minisec[1]{\textbf{#1}.}

\newcommand\Order{O}

\newcommand\del\partial
\newcommand\Tsym{\ensuremath{\textsf{T}}}
\newcommand\ZZ{\mathbb{Z}}

\newcommand\Dov{D_{\mathrm{ov}}}

\newcommand\Dbf{ \mathsf{D}}

\newcommand\Dbfov{\Dbf_{\mathrm{ov}}}

\newcommand\Rsym{\mathsf{R}}

\def\Vmaj{V_{\rm m}}
\def\Amaj{A_{\rm m}}

\newcommand\Dw{D_w}
\newcommand\Dcont{\mathcal{D}}
\newcommand\halfopen[1]{(#1]}

\usepackage[T1]{fontenc}

\title{Generalized Ginsparg-Wilson relations: Fermionic anomalies on the lattice}
\ShortTitle{Generalized Ginsparg-Wilson relations}

\author*[a]{Hersh Singh}

\affiliation[a]{Fermi National Accelerator Laboratory,\\
  Batavia, Illinois, 60510, USA}

\emailAdd{hershs@fnal.gov}

\abstract{
The \ac{GW} relation elegantly captures how the anomalous chiral symmetry of a Dirac fermion manifests on the lattice. In this talk, we discuss how the GW relation and its closed-form solution, the overlap operator, can be generalized to Majorana or Dirac fermions in any dimension for finite symmetry transformations (continuous or discrete). We find an exact symmetry which reproduces both perturbative and global anomalies on the lattice. These generalized GW fermions are boundary theories of various bulk symmetry-protected topological phases and thus provide an explicit lattice realization of the fermionic bulk-boundary correspondence central to recent proposals for chiral gauge theories on the lattice.
}

\FullConference{The 41st International Symposium on Lattice Field Theory (LATTICE2024)\\
  28 July - 3 August 2024\\
  Liverpool, UK\\}

\begin{document}
\maketitle

\glsresetall

\section{Introduction}

It has proven difficult to give a nonperturbative definition of the Standard Model.
This is due to our inability to formulate non-abelian chiral gauge theories on the lattice.
Despite many decades of effort, and even a solution for \emph{abelian} chiral gauge theories, the problem remains unsolved \cite{kaplan_chiral_2012, luscher_chiral_2002a, luscher_abelian_1999}.

The situation improves considerably when considering a \emph{global} chiral symmetry with an 't Hooft anomaly (as in quantum chromodynamics) rather than a gauged chiral symmetry (as in electroweak theory).
Developments related to the \ac{GW} relation \cite{ginsparg_remnant_1982} have clarified how an anomalous global chiral symmetry manifests optimally on the lattice \cite{luscher_exact_1998}, leading to a satisfactory resolution of the ``fermion-doubling'' problem \cite{kaplan_method_1992, neuberger_more_1998, neuberger_exactly_1998, hasenfratz_prospects_1998} and eventually even a nonperturbative definition for \emph{abelian} chiral gauge theories \cite{luscher_abelian_1999}.

Meanwhile, recent advances at the intersection of condensed-matter and particle physics have revealed a broader classification of fermionic anomalies \cite{kapustin_fermionic_2015, kitaev_periodic_2009a,  witten_fermion_2016, witten_anomaly_2020}. This includes the well-known perturbative and conventional global anomalies \cite{witten_su_1982a}, but also newer, more subtle global anomalies \cite{witten_fermion_2016, garcia-etxebarria_daifreed_2019, witten_anomaly_2020, witten_parity_2016, wang_new_2019}.
These new insights have emerged from the correspondence between \ac{SPT} phases in the bulk and 't Hooft anomalies on the boundary --- the so-called \emph{bulk-boundary} correspondence --- which
has led to a renaissance of attempts towards constructing
lattice chiral gauge theories
\cite{aoki_lattice_2024a, aoki_lattice_2024b, berkowitz_exact_2023, catterall_chiral_2021, catterall_lattice_2024, kaplan_chiral_2024, kaplan_weyl_2024, pedersen_reformulation_2023, wang_nonperturbative_2022, wang_solution_2019, wang_symmetric_2022a, zeng_symmetric_2022, golterman_conserved_2024, golterman_propagator_2024, clancy_generalized_2024, clancy_ginspargwilson_2024, butt_anomalies_2021}.

Given the importance of the \ac{GW} relation in understanding the chiral anomaly on the lattice, and the broadening of our understanding of fermionic anomalies, the central question of this talk is the following: can the \ac{GW} relation capture all fermionic anomalies? In other words, is there a \ac{GW} relation for any fermionic anomaly (or equivalently, an \ac{SPT} phase in one higher dimension)?

\begin{figure}[ht]
  \centering
  \includegraphics[width=0.7\linewidth]{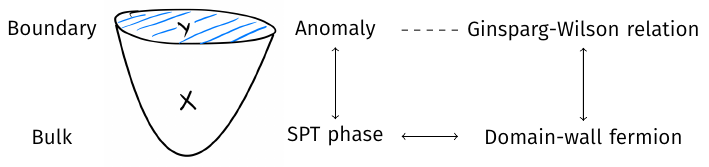}
  \caption{\label{fig:overview}
    The relation between conventional domain-wall fermions and the \ac{GW} relation for Dirac fermions is an explicit lattice realization of
    the general bulk-boundary correspondence between \ac{SPT} phases in the bulk and 't Hooft anomalies on the boundary.
    Are there \ac{GW} relations for all fermionic anomalies, or equivalently, all fermionic \ac{SPT} phases?
  }
\end{figure}

In this talk,\footnote{\emph{Notation:}
  Throughout this talk, we use $d$ for Euclidean spacetime dimensions. The Euclidean Dirac $\gamma_{\mu}$ ($\mu = 1,\dotsc, d$) matrices are chosen to be Hermitian so that the continuum Dirac operator $\slashed{\del} = \gamma^{\mu} \del_{\mu}$ is anti-Hermitian. We use the notation $D$ for a regulated Dirac operator and $\Dcont$ for a (unregulated) continuum Dirac operator.
  When using a lattice-regulated Dirac operator, we write the action simply as $S = \overline{\psi} D \psi$, suppressing the sum over any spatial, spinor or flavor indices.}
we discuss recent progress in using a \ac{GW} approach towards understanding fermionic anomalies on the lattice \cite{clancy_generalized_2024}.
We show that the original \ac{GW} construction, which applied to infinitesimal chiral symmetry for even-dimensional Dirac fermions, may be generalized to Majorana or Dirac fermions in arbitrary spacetime dimension and arbitrary (discrete or continuous) anomalous symmetry transformations.
We find that these generalized \ac{GW} relations apply to various bulk \ac{SPT} phases, reproducing the corresponding anomalies.

To show how the \ac{GW} relation can be generalized, we first recall the fermion doubling problem for Dirac fermions and how the standard \ac{GW} relation resolves it in \cref{sec:problem}.
Then, in \cref{sec:majorana}, we discuss generalizations of the \ac{GW} relation with the example of 1-dimensional Majorana fermions with a discrete time-reversal anomaly, following Ref.~\cite{clancy_generalized_2024}.
Finally, in \cref{sec:summary}, we comment on some open questions in this approach.

\section{Dirac fermions and the chiral anomaly}
\label{sec:problem}

Although the problem for a \emph{gauged} chiral symmetry is more severe,
there is in fact already a problem with the \emph{global} chiral symmetry for a Dirac fermion on the lattice.
Simply put, naive attempts to formulate Dirac fermions with an exact chiral symmetry on the lattice lead to additional unwanted fermions, called ``doublers''.
This can be seen by a straightforward discretization of the continuum Dirac action $S = \int\! d^{d}x\ \overline{\psi} \slashed{\del} \psi $, where
we simply replace the continuum derivative with a difference operator on a  $d$ (even)-dimensional hypercubic Euclidean spacetime lattice. In the absence of gauge fields, this becomes
\begin{align}
	\Dcont \psi = \gamma_{\mu} \del_{\mu} \psi \xrightarrow[\text{\quad discretization \quad}]{\text{naive}} D\psi
  = \frac{1}{2a} \textstyle\sum_{\mu}  \gamma_{\mu} (\delta_{\mu} + {\delta}^{*}_{\mu}) \psi
 = \frac{1}{2a} \textstyle\sum_{\mu} \gamma^{\mu} \left(\psi_{x + \hat{\mu}} - \psi_{x - \hat{\mu}}\right)
  \label{eq:naive}
\end{align}
where $a$ is the lattice spacing,
$\gamma_{\mu}$ are the (Hermitian) Dirac matrices,
and $\delta_{\mu}, {\delta}^{*}_{\mu}$ are forward and backward difference operators, respectively, acting on lattice Dirac fields $\psi_{x}$ as
$ \delta_{\mu} \psi_{x} =   \psi_{x + \hat{\mu}} - \psi_{x}$ and ${\delta}^{*}_{\mu} \psi_{x} = \psi_x - \psi_{x - \hat{\mu}} $,  with
$\hat{\mu}$ being a unit vector in the direction $\mu=1,\dotsc, d$.
In momentum space, the continuum Dirac operator $\Dcont(p) = \gamma_{\mu} p_{\mu}$ becomes the lattice Dirac operator $D(p) = \frac{i}{a} \sum_{\mu} \gamma^{\mu} \sin( a p_{\mu} )$, where $a p_{\mu} \in \halfopen{-\pi, \pi}$ is now a periodic variable.
We see that $D(p)$ has $2^{d}$ zeros, which result in propagator poles.
These are additional massless fermions which survive the continuum limit, but are clearly not present in the continuum theory we started with.
This is the fermion doubling problem.
At this point, though, the problem seems to depend on our choice of the naive discretization scheme in \cref{eq:naive}.
So we may wonder: is there a clever discretization?

This is where the no-go theorem of \ac{NN} \cite{nielsen_nogo_1981, nielsen_absence_1981a, nielsen_absence_1981} comes in.
The \ac{NN} no-go theorem shows that a local \emph{free}-fermion lattice theory with the right continuum limit and an exact chiral symmetry cannot be free of doublers.
Therefore, we must ask a sharper question: which of the conditions of the no-go theorem should we give up to satisfactorily formulate a single massless Dirac fermion on the lattice?

As an example of how to remove the doublers, we can add a momentum-dependent mass term to the naive discretization of \cref{eq:naive}.
The mass term is chosen so that the doublers at the corners of the Brillouin zone become heavy, while the fermion at the origin stays massless.
This can be achieved with the Wilson-Dirac operator,
\begin{align}
	\Dw(m,r) &= \sum_{\mu} \frac{1}{2a} \left[ \gamma_{\mu} (\delta_{\mu} + \delta_{\mu}^{*}) - r \delta^{*}_{\mu} \delta_{\mu} \right] + m
        \label{eq:Dw}
\end{align}
where $m$ is a bare fermion mass, $r$ is called the Wilson parameter, and
$\Delta = {\delta}^{*}_{\mu} \delta_{\mu}$ is a lattice discretization of the Laplacian operator.
In momentum space, this operator is
$ \Dw(p) =  \sum_{\mu} \frac{1}{a} \left[ i \gamma_{\mu} \sin(a p_{\mu}) + r (1 - \cos a p_{\mu}) \right] + m $.
With $m=0$ and $r \sim \Order(1) > 0$, all the doublers acquire a heavy mass of cutoff scale $\sim 1 / a$, while the fermion at the origin stays massless.
The problem with this approach is that while it removes the doublers, it unfortunately also breaks chiral symmetry since $\{D_w, \gamma_5 \} \neq 0$.
The chiral symmetry is only recovered in the continuum limit.
This becomes a practical problem when we turn on gauge fields, because in the absence of chiral symmetry, fermion masses get additively renormalized and require fine tuning.
Thus, we sacrifice chiral symmetry to eliminate the doublers, consistent with the \ac{NN} no-go theorem.
There are, of course, numerous other ways to violate one of the conditions of the \ac{NN} theorem, but is there an ``optimal'' approach?

\minisec{The Ginsparg-Wilson relation}
To understand the nature of chiral symmetry on the lattice, Ginsparg and Wilson \cite{ginsparg_remnant_1982} asked the question: what becomes of the exact chiral symmetry when a continuum theory of massless Dirac fermion with an exact chiral symmetry $\{\Dcont , \gamma_5\} = 0$ undergoes a \ac{RG} ``block-spin'' transformation, leading to a lattice theory of the block-averaged lattice fields?
Denoting the resulting lattice Dirac operator by $D$, \ac{GW} found that it satisfies a remarkable relation
\begin{align}
  \{\Dcont, \gamma_5\} = 0 \xrightarrow{\text{\ac{RG} blocking}} \{ D, \gamma_5\} = 2 a D \gamma_{5} D,
  \label{eq:gw}
\end{align}
where $a$ is the lattice spacing.
This is the \ac{GW} relation.
It reveals that the lattice Dirac operator violates chiral symmetry, but in a rather specific manner.
Despite formulating this relation, \ac{GW} could not find explicit solutions to it, leading their work to be largely overlooked for years.
The relation was later rediscovered when Hasenfratz serendipitously found that perfect actions satisfy the \ac{GW} relation \cite{hasenfratz_prospects_1998}.
Following this, it was quickly realized that overlap \cite{neuberger_more_1998, neuberger_exactly_1998}
and domain-wall fermions \cite{kaplan_method_1992} are also solutions.
Finally, Lüscher showed that the \ac{GW} relation implies an \emph{exact}, but modified, chiral symmetry for any Dirac operator which satisfies it \cite{luscher_exact_1998}.
This cemented the fundamental role of the \ac{GW} relation in understanding lattice chiral symmetry and explaining the remarkable chiral properties of its various solutions.\footnote{See also the talk by Jan Smit in this conference \cite{smit_confederacy_2025} for an interesting account of the field \emph{before} the developments related to the \ac{GW} relation.}

\minisec{The overlap operator and Lüscher symmetry}
The overlap operator, in particular, provides an elegant closed-form solution to the \ac{GW} relation \eqref{eq:gw}.
In terms of the Wilson-Dirac operator $\Dw$ [\cref{eq:Dw}], the overlap operator $\Dov$ is given by
\begin{align}
  a\Dov &= \frac{1}{2} (1 + V), \quad V = A / {\left[{A}^{\dagger} A \right]}^{-\frac{1}{2}}.
\end{align}
where $A = \Dw(-m, am)$ is a Wilson-Dirac operator [\cref{eq:Dw}] with $m > 0$ and the Wilson coupling $r = ma$.
It is a straightforward exercise to check that the overlap operator $\Dov$ satisfies the \ac{GW} relation, is free of doublers, and has the right continuum limit.
Unlike the Wilson-Dirac operator, the overlap operator has a remarkable property by the virtue of being a solution to the \ac{GW} relation.
In fact, for any $D$ satisfying the \ac{GW} relation, the lattice action $S = \overline{\psi} D \psi $ has an \emph{exact} chiral symmetry, albeit an unusual one \cite{luscher_exact_1998}:
\begin{align}
  \psi &\to \psi( 1 - i \varepsilon \gamma_{5} V) , \quad
      \overline{\psi} \to \overline{\psi}(1 + i \varepsilon \gamma_{5})
  \label{eq:luscher-sym}
\end{align}
Note that the presence of $V$ in the transformation for $\psi$ makes this transformation non-onsite.
The continuum limit arises when $V \to -1$, so this reduces to the ordinary chiral transformation in that limit.\footnote{Note that $\psi$ and $\overline{\psi}$ transform differently --- this is only possible in Euclidean spacetime.
We shall return to this point later when we look at Majorana fermions.}
This symmetry protects the masslessness of Dirac fermions against additive mass renormalization, thus eliminating the fine-tuning problem that plagues the Wilson-Dirac formulation.

\minisec{Anomaly on the lattice}
We have seen that the overlap action has an exact (modified) chiral symmetry.
But recall that the $U(1)_{\chi}$ chiral symmetry has a mixed anomaly with the $U(1)_{V}$ vector symmetry.
That is, in the presence of background $U(1)_{V}$ gauge fields, the $U(1)_{\chi}$ chiral symmetry is broken.
If the lattice action is invariant under the modified chiral symmetry of \cref{eq:luscher-sym}, how does this anomaly emerge?
It turns out that the \ac{GW} fermions reproduce the anomaly precisely as continuum fermions do --- via the noninvariance of the fermionic path integral measure \cite{fujikawa_pathintegral_1979}.
The Jacobian associated with transformation \cref{eq:luscher-sym} is
$e^{-i \varepsilon \tr \gamma_5 V}$
where $\tr \gamma_5 V = \operatorname{ind} \Dov$ is in fact the index of the overlap Dirac operator, defined as the difference between the number of negative and positive chirality zero modes. This is precisely the expected anomaly.

\minisec{A continuum solution to the \ac{GW} relation}
Much intuition can be gained about the overlap operator, and its generalizations which we will discuss in the next section, by examining a ``continuum'' solution to the \ac{GW} relation.
Let us write a (regulated) Dirac operator as $D = h/(1+h) $,
where $h$ is some operator.
Substituting this into the \ac{GW} relation (with lattice spacing $a=1$) yields $\{ h, \gamma_5\} =0 $.
Thus, any operator $h$ with exact chiral symmetry provides a solution.
One natural choice is the continuum Dirac operator $\slashed{D} / m$ itself, where $m$ is a regulator mass scale. This yields a continuum solution to the \ac{GW} relation:
$ D = {\slashed{D}}/{ (\slashed{D} + m) }. $
  But this is just a \ac{PV} regulated Dirac fermion! A \ac{PV} regularization will produce this Dirac operator upon integrating out the massive, complex, bosonic ghost field $\phi$ with the action $S_{\text{ghost}} = \int\! d^d\!x\ \overline{\phi} (\slashed{D} + m) \phi$.
  Therefore, we find that a continuum \ac{PV} regulated fermion already satisfies the \ac{GW} relation.
In a sense, the overlap operator is a nonperturbative lattice version of the perturbative continuum \ac{PV} regularization.

Interestingly, the \ac{PV} solution to the \ac{GW} relation also illuminates the origin of the modified chiral symmetry in \cref{eq:luscher-sym}.
While the transformation in \cref{eq:luscher-sym} looks mysterious for the overlap operator, it has simple interpretation for the \ac{PV} regulated fermion: it is just the transformation under which the fermion field transforms while the ghost bosonic fields do not \cite{clancy_generalized_2024}.

\section{Majorana fermions and a time-reversal anomaly}
\label{sec:majorana}

\begin{figure}[ht]
  \centering
  \includegraphics[width=0.9\linewidth]{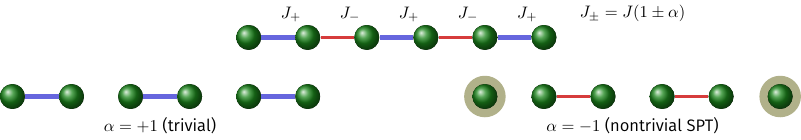}
  \caption{\label{fig:kitaev}
    Fidkowski-Kitaev chain with bond staggering $\alpha$ on an open chain with an even number of lattice sites.
    The system is critical at $\alpha = 0$.
    For $\alpha > 0$, the system enters a trivial gapped phase, while $\alpha < 0$ corresponds to a nontrivial \ac{SPT} phase.
    The \ac{SPT} phase ($\alpha < 0$) hosts edge modes which are (0+1)-dimensional Majorana fermions protected by an anomalous time-reversal symmetry $\Tsym$ with $\Tsym^2=+1$.
    Our goal is to obtain a \ac{GW} formulation of the Majorana edge modes and explicitly reproduce the time-reversal anomaly.
  }
\end{figure}

We have seen how the \ac{GW} relation and its solution, the overlap operator, provide an elegant lattice framework which captures the 't Hooft anomaly involving the chiral and vector $U(1)$ symmetries.
In this section, we illustrate a generalization of the \ac{GW} relation by examining Majorana fermions in odd dimensions with a global discrete anomaly.

To motivate the anomaly, let us consider the Fidkowski-Kitaev chain. This is a 1+1-dimensional Hamiltonian model of single-component Majorana fermions on an $L$-site lattice with a time-reversal symmetry $\Tsym$ satisfying $\Tsym^{2} = +1$.
Let $\lambda_j$ denote single-component Majorana fermions at sites $j=1,\dotsc, L$ satisfying the anti-commutation relations $\{\lambda_i, \lambda_{j}\} = \delta_{ij}$. For an open chain with even $L$, the Hamiltonian takes the form:
$
  H = J \sum_{j=1}^{L} (1 + \alpha (-1)^{j}  )\lambda_j \lambda_{j+1},
$
where $J_{\pm} = J (1 \pm \alpha)$ is a staggered coupling for even and odd links, as illustrated in \cref{fig:kitaev}.
Our interest in this system lies in the fact that it exhibits a nontrivial \ac{SPT} phase and corresponding edge modes.

This system has a second-order critical point at $\alpha = 0$ and remains gapped for any $\alpha \neq 0 $.
However, this is an important distinction between $\alpha > 0$ and $\alpha < 0$: the former is a trivial phase, while the latter is a nontrivial \ac{SPT} phase.
This can be seen by considering the limits $\alpha = \pm 1$ on an open chain, shown in \cref{fig:kitaev}.
At $\alpha = +1$, all neighboring Majorana fermions couple, resulting in a trivial gapped phase.
In contrast, at $\alpha = -1$, the Majorana modes at the edges ($j=1, L$) are completely decoupled from the rest of the chain, while the remaining Majorana modes are paired.
The decoupled Majoranas at sites $j=1,L$ together form a $2$-dimensional Hilbert space, giving rise to 2-dimensional ground state degeneracy.
This degeneracy is robust against any $\Tsym$-preserving perturbations \cite{fidkowski_topological_2011, fidkowski_effects_2010}. Such robust edge-modes are the hallmark of a nontrivial \ac{SPT} phase.

The underlying physics here is analogous to the anomaly-inflow mechanism \cite{callan_anomalies_1985} which provides the foundation for domain-wall fermions \cite{kaplan_method_1992}.
Indeed, this \emph{bulk-boundary correspondence} is quite general and establishes a one-to-one correspondence between \ac{SPT} phases and 't Hooft anomalies.
Physically, this means that any system in a nontrivial \ac{SPT} phase necessarily hosts corresponding edge-modes at its boundary.
Both domain-wall fermions and the Majorana chain discussed earlier are examples of this.
In this language, the 5-dimensional domain-wall fermion realizes a nontrivial \ac{SPT} phase.
This is why we get robust massless Weyl fermions at the 4-dimensional boundary.

The edge mode of the 1+1-dimensional Majorana chain is simply a 0+1-dimensional Majorana fermion, with the continuum action $ S = \int\! dt\, \chi \del_{t} \chi, $
where $\chi$ is a single-component boundary Majorana fermion.
This appears deceptively simple --- can such a system have an anomaly?
Indeed, as explained by Witten~\cite{witten_fermion_2016}\footnote{See also Witten-Yonekura \cite{witten_anomaly_2020} for a more thorough discussion of the general fermionic bulk-boundary correspondence.}, this system has a subtle (global) $\ZZ_8$ anomaly involving time-reversal and fermion parity symmetries.
The question for us is thus: can we recover this anomaly on the lattice within the \ac{GW}/overlap framework?\footnote{For an illuminating discussion of this anomaly in the Hamiltonian framework, see Ref.~\cite{witten_anomalies_2023} and also Ref.~\cite{seiberg_majorana_2024} for a related discussion in 1+1 dimensions. Here, we are interested in a Euclidean lattice perspective on this anomaly.}

\minisec{Generalizing the GW relation to Majorana fermions}
In Ref.~\cite{clancy_generalized_2024}, it was shown that the original derivation of \ac{GW} can be generalized beyond even-dimensional Dirac fermions and infinitesimal (continuous) symmetries to include odd dimensions, Majorana fermions, and finite symmetry transformations (including discrete symmetries).
We refer the reader to the paper for details.
Here, we sketch the essential features of the procedure and highlight some subtleties.

When employing the \ac{GW} construction,
we need to add a Gaussian term to the original continuum action, coupling the continuum fields to the block-averaged lattice fields.
Upon integrating out the original continuum fields, one obtains a lattice theory.
This Gaussian term has the dimensions of a mass term, which is how a regulator scale enters the problem and how the standard (or continuum) version of chiral symmetry (corresponding to $\{\Dcont, \gamma_{5 } \} = 0$) is broken by the regulator.

A mass term is easy to write down for a Dirac fermion.
But Fermi statistics prohibit adding such a term to the Majorana action $S = \int\! dt\, \chi \del_{t} \chi$.
Consequently, we cannot formulate a \ac{GW} relation for a single 0+1-dimensional Majorana fermion.\footnote{This, incidentally, is the same problem with formulating a \ac{GW} relation for a single Weyl fermion, which also does not allow a mass term.}
This becomes possible, however, for two Majorana fermions, where one can introduce a mass term of the form $\chi_{a}\varepsilon_{ab} \chi_{b}  $, with flavor indices $a,b=1,2$ and $\varepsilon_{ab}$ being the totally antisymmetric Levi-Civita tensor.
The continuum action with a mass term thus takes the form
$ S = \int\!dt\, \chi^{T} (\del_{t} + \mu \tau_{2}) \chi, $
where $\chi$ is now a two-flavor one-dimensional Majorana field,
and $\tau_{2}$ is the second Pauli matrix.

Reflection symmetry $\Rsym$, which is the same as time-reversal in Euclidean spacetime \cite{witten_fermion_2016}, can be taken to act on the Majorana fields as
$
	\chi(t) \to \Rsym \chi(t) = i \chi(-t),
$
such that $\Rsym^{2} = -1$.
One can verify that this transformation preserves the kinetic term but not the mass term.
This is a hint that this symmetry might be anomalous.
At this point, we have all the ingredients to apply the \ac{GW} construction, as detailed in Ref.~\cite{clancy_generalized_2024}.
Summarizing the results:
if $ S = \chi^{T} \Dbf \chi $ is the \ac{GW} Majorana lattice action with an antisymmetric lattice Majorana-Dirac operator $\Dbf$, then we find the relation
\begin{align}
  \Dbf_{\Rsym} - \Dbf = 2 a \Dbf \tau_{2} \Dbf_{\Rsym},
\end{align}
where $\Dbf_{\Rsym} = \Rsym^{T} \Dbf \Rsym $.
This is the \ac{GW} relation for a 2-flavor Majorana system in $d=1$ corresponding to the reflection symmetry $\Rsym$.
Importantly, an explicit overlap solution exists:
\begin{align}
	a \Dbfov &= \frac{\tau_{2}}{2} (1 + \Vmaj), \quad
  \Vmaj = \Amaj / [\Amaj^{\dagger} \Amaj ]^{-\frac{1}{2}},
\end{align}
where $\Amaj = \Dw^{\text{m}}(-\mu, a \mu)$ and $ \Dw^{\text{m}}(\mu,r) = \frac{1}{2a} [ \tau_2\ (\delta_{1} + \delta_{1}^{*}) - r \delta_1^{*} \delta_{1}] + \mu$ is the analog of the Wilson operator [\cref{eq:Dw}] for the 2-flavor Majorana fermion in $d=1$ with $\mu > 0$.

\minisec{Lüscher symmetry for Majorana fermions}
Does the \ac{GW} Majorana action admit an exact modified reflection symmetry, analogous to the modified chiral symmetry for Dirac fermions?
At the outset, we note that a transformation like \cref{eq:luscher-sym} cannot work, because we do not have the freedom of transforming $\overline{\psi}$ and $\psi$ separately for Majorana fermions.  Nevertheless, a viable symmetry transformation exists \cite{clancy_generalized_2024}:
$ \chi \to \Rsym \sqrt{-\Vmaj} \chi. $
The definition of this square root requires care.
With the appropriate definition, we find again the possibility of a nontrivial Jacobian for this symmetry transformation.
The Jacobian evaluates to $\det( \Rsym \sqrt{-\Vmaj} ) = (-1)^{\frac{\nu_{-}}{2}}$, where $\nu_{-}$ counts the number of zero modes of $\Dbf$ (equivalently, the number of $\Vmaj = -1$ modes).
This is precisely the mod-2 index of the Dirac operator which appears in the continuum anomaly.
Since we find a $\ZZ_{2}$ anomaly for the 2-flavor Majorana, this means there is (at least) a $\ZZ_{4}$ anomaly for a single $d=1$ Majorana in the \ac{GW} framework.
Interestingly, the complete anomaly is known to be $\ZZ_{8}$ \cite{fidkowski_topological_2011, kapustin_fermionic_2015, witten_fermion_2016}, suggesting limitations in our current framework.

\section{Conclusions}
\label{sec:summary}

We have shown how the original \ac{GW} relation extends beyond the chiral anomaly for Dirac fermions to encompass odd dimensions, discrete symmetries, and Majorana fermions.
As a concrete example, we examined the Fidkowski-Kitaev chain, which hosts (0+1)-dimensional Majorana edge modes with a $\ZZ_{8}$ anomaly involving time-reversal and fermion parity.
\Cref{tab:summary} summarizes these results.
While previous work has explored various generalizations of the \ac{GW} relation \cite{bietenholz_ginspargwilson_2001, so_ginsparg_1999, kimura_band_2023a, kikukawa_overlap_1998, cundy_ginspargwilson_2011, bergner_generalising_2009a, bietenholz_exact_1999, bietenholz_solutions_1999}, we present here a unified treatment that derives a \ac{GW} relation in all cases where a continuum \ac{PV} regularization exists.

\begin{table}\smaller
  \centering
  \renewcommand{\arraystretch}{1.1}
  \begin{tabular}{l | c | c}
    \TopRule
    & \emph{Dirac, $d=4$} & \emph{Majorana, $d=1$} \\
    \MidRule
    Continuum action  & $S = \int\! d^{4}x\ \bar \psi ( \slashed{D} + m) \psi $ & $S = \int\! dt\ \  \chi^T (\del_{t}  + m \tau_{2} ) \chi $ \\
    Symmetry & Chiral symmetry (continuous)
                          & Time-reversal or Reflection (discrete) \\
    Anomaly & Perturbative & Global \\
    \ac{GW}  relation
    & $ \{ \gamma_5, D\} = 2a D \gamma^5 D $
                          & $\Dbf_{\Rsym} - \Dbf = 2 a \Dbf \tau_{2} \Dbf_{\Rsym}$
    \\
    Lattice symmetry & $\delta \psi = - \gamma_{5} V \psi, \quad \delta\bar \psi = \bar \psi \gamma_5 $ & $\chi \to R \sqrt{-\Vmaj}\ \chi$
    \\
    Jacobian & $ e^{-i \varepsilon\ \mathrm{ind} D  }$ & $(-1)^{\frac{\nu_{-}}{2}}$ \\
    Topological invariant & index  & mod-2 index \\
    \BotRule
  \end{tabular}

  \caption{\ac{GW} formulation for Dirac fermions in $d=4$ and Majorana fermions in $d=1$ dimensions.}
  \label{tab:summary}
\end{table}

The connection between \ac{SPT} phases in the bulk and 't Hooft anomalies on the boundaries has proven remarkably fruitful over the past few decades.
From this point of view, domain-wall fermions constitute a bulk system in an \ac{SPT} phase, whereas the overlap operator and the \ac{GW} relation furnish an explicit lattice realization of the boundary theory with an exact 't Hooft anomaly.
In this work, we investigated whether a \ac{GW} relation exists for all fermionic anomalies (and consequently for all fermionic \ac{SPT} phases).
We found that while the \ac{GW} construction admits substantial generalization, certain crucial elements remain missing.
For example, it is unclear how the complete $\ZZ_{8}$ Dai-Freed anomaly for the 1-dimensional Majorana emerges in this formulation, which only makes a $\ZZ_{4}$ anomaly manifest.
A satisfactory Hamiltonian formulation remains another open question --- one that would be crucial for quantum computing applications and would help bridge the divide between this approach and the condensed matter literature.
Given the prominent role of the bulk-boundary correspondence in recent approaches to constructing lattice chiral gauge theories, one is led to wonder:
Could a complete picture of lattice fermionic anomalies be the missing piece needed to solve the longstanding puzzle of non-abelian chiral gauge theories?
This remains to be seen.

\acknowledgments
I thank Michael Clancy and David Kaplan for our collaboration on Ref.~\cite{clancy_generalized_2024}, which forms the basis of this talk.
I am grateful to Wolfgang Bietenholz, Claude, Hank Lamm, Mendel Nguyen, and for their valuable comments on the manuscript.
I also thank the organizers of Lattice '24 for an outstanding conference, and the participants --- especially Evan Berkowitz, Simon Catterall, Aleksey Cherman, George Fleming, Hidenori Fukaya, Tetsuya Onogi, Srimoyee Sen, and David Tong --- for illuminating discussions on related issues.
This work is supported by the Department of Energy through the Fermilab QuantiSED program in the area of ``Intersections of QIS and Theoretical Particle Physics.''
This manuscript has been authored by Fermi Forward Discovery Group, LLC under Contract No. 89243024CSC000002 with the U.S. Department of Energy, Office of Science, Office of High Energy Physics.

\bibliographystyle{JHEP}
\bibliography{refs}

\end{document}